\documentclass[12pt,a4paper]{article}
\usepackage{a4}
\usepackage[english]{babel}
\usepackage{cite}
\usepackage{graphicx}
\newcommand{\EE}{\rm e^+ e^-}
\begin{document}
\begin{titlepage}
\begin{flushright}
DESY 06-065\\
9th May 2007
\end{flushright}
\vspace*{3cm}
\begin{center}
\boldmath
\textbf{\Large 
Towards a measurement of the two-photon decay width of the Higgs boson at a 
Photon Collider 
}
\end{center}

\begin{center}
\textbf{K. M\"onig$^1$ and A. Rosca$^2$}
\end{center}

$^1$ DESY, Zeuthen, D 15738, Germany
 
$^2$ West University of Timisoara, Timisoara, RO 300223, Roumania
 
%
%
\begin{abstract}
  A study of the measurement of the two photon decay width times the
  branching ratio of a Higgs boson with the mass of 120 GeV in photon
  - photon collisions is presented, assuming a $\gamma \gamma$
  integrated luminosity of 80 fb$^{-1}$ in the high energy part of the
  spectrum. The analysis is based on the reconstruction of the Higgs
  events produced in the $\gamma \gamma \to {\rm H}$ process, followed
  by the decay of the Higgs into a ${\rm b}\bar{\rm b}$ pair. A
  statistical error of the measurement of the two-photon width,$\Gamma
  (H \to \gamma \gamma)$, times the branching ratio of the Higgs
  boson, BR$(H \to {\rm b}\bar{\rm b})$ is found to be 2.1 $\%$ for
  one year of data taking.
\end{abstract}
%
%

\end{titlepage}

\section{Introduction}

The central challenge for particle physics nowadays is the origin of
mass. In the Standard Model the masses of both fermions and gauge
boson are generated through interactions with the same scalar
particle, the Higgs boson H. While it can only be produced in
association with another particle at an ${\rm e}^{+}{\rm e}^{-}$
collider, the Higgs boson can be produced singly in the s-channel of
the colliding photons at a Photon Collider.  If it exists, the Higgs
boson will certainly be discovered by the time such a facility will be
constructed. The aim of this machine will be then a precise
measurement of the Higgs properties. This facility as an upgrade
option at the ILC \cite{scope} will permit a high precision
measurement of the ${\rm H} \to \gamma \gamma$ partial width, which is
a quantity sensitive to the existence of new charged particles.  For
this reason such a measurement is significantly important.  If we find
a deviation of the two photon width from the Standard Model prediction
it means that an additional contribution from unknown particles is
present, and so it is a signature of physics beyond the Standard
Model. For example, the minimal extension of the Standard Model
predicts the ratio of the two photon width $\Gamma(\rm H \to \gamma
\gamma, \rm MSSM)$/ $\Gamma(\rm H \to \gamma \gamma, \rm SM) < $ 1.2
\cite{mssm} for a Higgs boson with a mass of 120 GeV, assuming a
supersymmetry scale of 1 TeV and the chargino mass parameters $M$ and
$\mu$ of 300 and 100 GeV, respectively.

At a Photon Collider one can measure the product
$\Gamma(\rm H \to \gamma \gamma)$$\times$BR$(\rm H \to \rm X)$. To 
obtain the two-photon partial width independent of the
branching ratio one has to combine the above measurement with an accurate
measurement of the BR($\rm H \to \rm X$) from another
machine.

This study investigates the capability of an ILC detector to measure the two
photon decay width times the branching ratio
for a Higgs boson with the mass of 120 GeV, the
preferred mass region by recent electroweak data \cite{ew}.
The simulation of the signal and background processes is described in section 2.
Event selection is described in section 3.
Results are summarised in section 4.

The feasibility of the measurement of the two photon decay width of the
Higgs boson in
this mass region has also been reported by [4-6]. Our analysis presents 
for the first time a realistic simulation of the background processes,
particularly the emission of a hard gluon. 

\section{Simulation of the signal and background processes}

The cross section for the Higgs boson formation is given by a
Breit-Wigner approximation $$\sigma_{\gamma \gamma \to \rm H}=8 \pi
\frac{\Gamma(\rm H \to \gamma \gamma) \Gamma_{\rm tot}}{(s _{\gamma
    \gamma}-M_{\rm H}^{2})^{2}+M^{2}_{\rm H} \Gamma^{2}_{\rm
    tot}}(1+\lambda_{1}\lambda_{2}),$$
where $M_{\rm H}$ is the Higgs
boson mass, $\Gamma(\rm H \to \gamma \gamma)$ and $\Gamma_{\rm tot}$
are the two photon and total decay width of the Higgs boson,
$\lambda_{1}$ and $\lambda_{2}$ are the initial photon helicities and
$\sqrt s_{\gamma \gamma}$ is the $\gamma \gamma$ centre-of-mass
energy.  The initial photons should have equal helicities, so that
$J_{\rm z}$ = 0, in order to make a spin-0 resonance as it is the case
of the Higgs boson. If polarised photon beams are used, the signal
cross section is increased up to a factor of 2.  The experimentally
observed cross section is obtained by folding this basic cross section
with the $\gamma \gamma$ collider luminosity distribution.

A Higgs boson with standard model coupling and a mass of 120 GeV
can be produced in the $\gamma \gamma \to$ H
process. In this mass region the Higgs particle will
decay dominantly into a b${\rm \bar{b}}$ pair. The event rate is given by the formula:
$$N(\gamma \gamma \to \rm H \to \rm b \bar{\rm b})=
\frac{d{\cal L}_{\gamma \gamma}}{\rm d \sqrt
s _{\gamma \gamma}}|_{M_{\rm H}} \frac{4 \pi^{2}\Gamma (\rm H \to \gamma \gamma)
\rm BR (\rm H \to \rm b \bar{\rm b})}{M_{\rm H} ^{2}} (1+\lambda_{1}\lambda_{2})(\hbar c)^{2},$$
where the conversion factor $(\hbar c)^{2}$ is  3.8937966$\cdot$$10^{11}$ fb GeV$^{2}$.
This rate depends strongly on the value of
the differential luminosity at the Higgs mass, 
$\frac{d{\cal L}_{\gamma \gamma}}{\rm d \sqrt
s _{\gamma \gamma}}|_{M_{\rm H}}$. 

High energy photon beams can be produced at
a high rate in Compton backscattering of laser photons off
high energy electrons \cite{pc}. 
The beam spectra at $\sqrt s_{ee}$ = 210 GeV are simulated
using the CompAZ \cite{compaz}, a fast parameterisation which includes
multiple interactions and non-linearity effects.
The shape of the luminosity distribution depends on the
electron and laser beam parameters. The electron and laser beam energy
considered for this study are 105 GeV and
1 eV, respectively, resulting in the maximum photon energy of about 70
GeV, suitable to study a Higgs boson with the mass of 120 GeV.
Setting opposite helicities for
the laser photons and the beam electrons the energy
spectrum of the backscattered photons is peaked at about 60$\%$
of the ${\rm e^{-}}$ beam energy. 
The number of high energy scattered photon is nearly two times higher if we
use polarised photons and electrons with opposite helicities than in the case of
unpolarised electron and laser photons. Consequently, this leads to an
improved luminosity in the high energy part of the spectrum.
The scattered photons
are highly polarised in this high energy region.
The helicity combination
of the two high energy photons can be arranged such that $J_{\rm z}$ = 0 state
is dominant.
The resulting value of $\frac{d{\cal L}_{\gamma \gamma}}{\rm d \sqrt
s _{\gamma \gamma}}|_{M_{\rm H}}$ is 1.6 fb$^{-1}$/GeV in one year of running 
using the parameters from \cite{pc}.

The branching ratios BR(H $\to \gamma \gamma$), BR(H $\to \rm b
\bar{\rm b}$) and the total width are taken to be 0.22$\%$, 68$\%$ and
4 MeV, respectively. These numbers are calculated with HDECAY
\cite{hdecay} program and include QCD radiative corrections.  With an
integrated luminosity of 80 fb$^{-1}$ per year in the hard part of the
spectrum \cite{pc} about 20000 signal events can be produced under
these conditions.

The signal $\gamma \gamma
\to {\rm H} \to {\rm b} \bar{\rm b}$ process is simulated with PYTHIA
\cite{pythia}. A total of 100K events were generated.
Parton evolution and hadronisation are simulated using
the parton shower and the string fragmentation models.

\begin{figure*}[t]
\centering
\includegraphics[width=65mm]{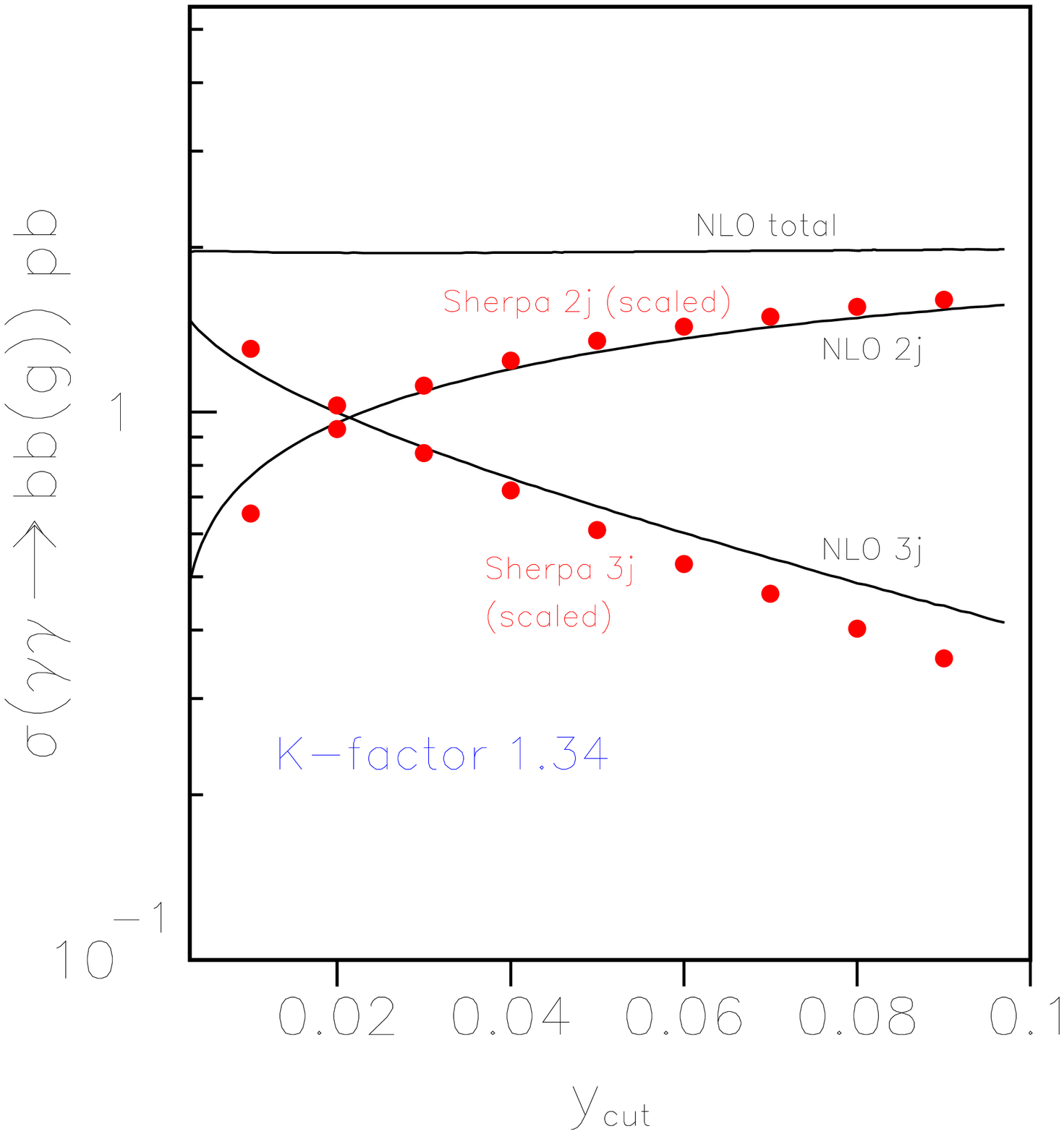}
\includegraphics[width=65mm]{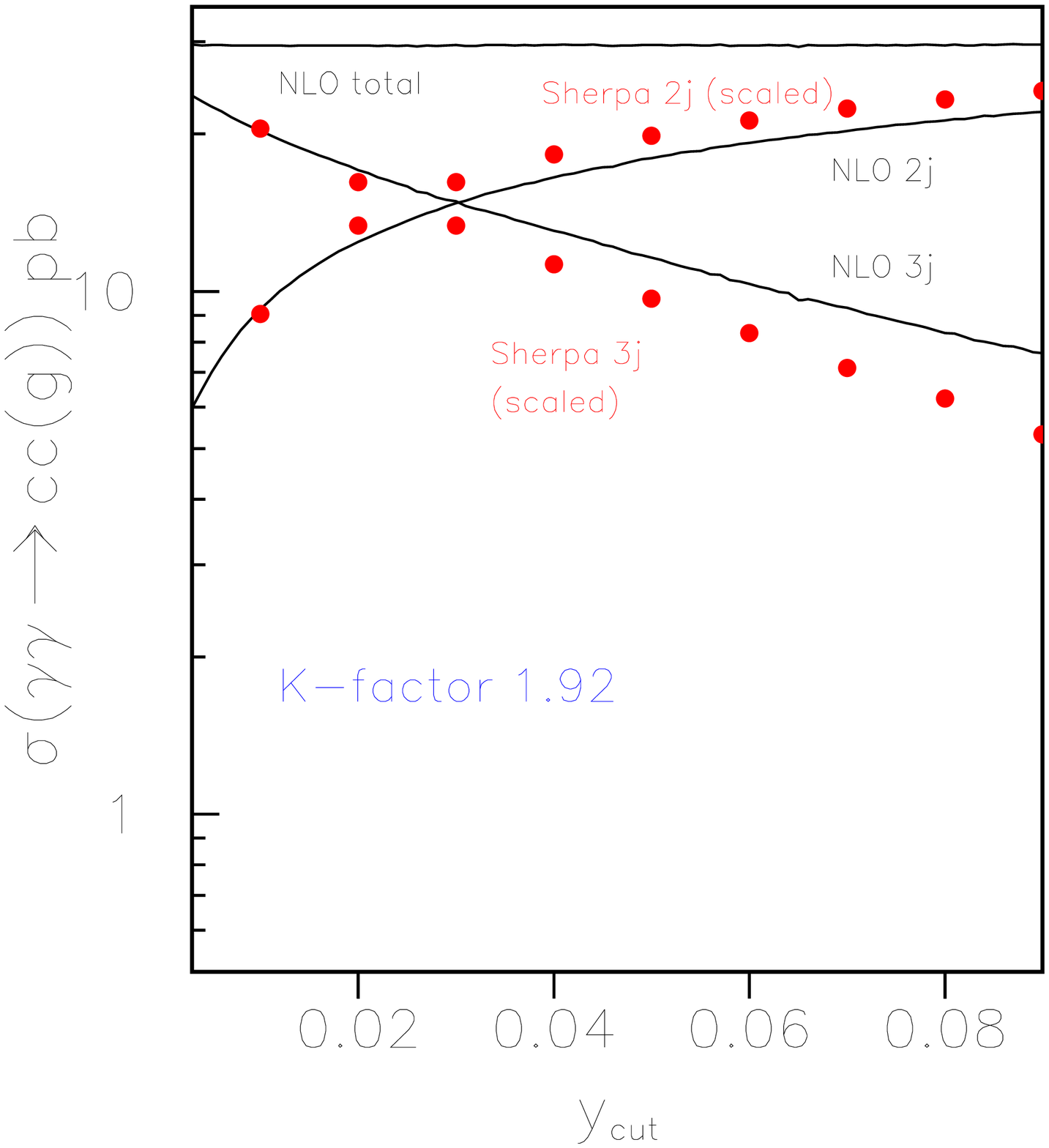} \\
\caption{Scaled SHERPA cross sections for $J_{z}$ = 0 compared to the NLO calculations
for a) $\gamma \gamma \to b\bar{b}(g)$ and b) $\gamma \gamma \to c\bar{c}(g)$.
}
\label{sh-nlo}
\end{figure*}

The main background processes to an intermediate mass Standard Model
Higgs production are the direct continuum $\gamma \gamma \to \rm b
\bar{\rm b}$ and $\gamma \gamma \to \rm c \bar{\rm c}$ production. The
light quarks are very efficiently rejected by the b-tagging.  Due to
helicity conservation, the continuum background production proceeds
mainly through states of opposite photon helicities, making the states
$J_{\rm z} = 2$. Choosing equal helicity photon polarisations the
cross section of the continuum background is suppressed by a factor
$M_{\rm q}^{2}/s_{\gamma \gamma}$, with $M_{\rm q}$ being the quark
mass. Unfortunately, this suppression does not apply to the process
$\gamma \gamma \to \rm q \bar{\rm q} \rm g$, because after the gluon
radiation the $\rm q \bar{\rm q}$ system is not necessarily in a
$J_{\rm z} = 0$ state. The resulting background is still very large
compared to the signal.  Therefore, a reliable prediction of the
background implies to consider the NLO QCD corrections.  Exact
one-loop QCD corrections have been calculated in \cite{qcd1} for both
$J_{z}$ = 0 and $J_{z}$ = 2 states and most recently in \cite{qcd2}.
For $J_{z}$ = 0 state it has been found that double logarithmic
corrections are also necessary and these were calculated and resumed
to all orders in the form of a non-Sudakov form factor in \cite{qcd3}.
For the background studies the SHERPA \cite{she} generator has been
used. SHERPA is a tree level matrix element generator which uses the
CKKW \cite{ckkw} method to merge the matrix elements for parton
production with the parton shower.  Using a jet algorithm, the
kinematic range for $n$ partons is partitioned into two regions, a
region of jet production which is covered by the corresponding matrix
elements, and a region of jet evolution which is covered by the parton
shower. In the matrix element dominated region the hard kinematics is
that of $n$ partons while in the parton shower dominated region the
hard kinematics is that relevant to $n-1$ partons. In both regions,
the matrix elements are reweighted with a combination of Sudakov form
factors entering the shower algorithm.  The hard emissions in the
parton shower leading to a jet are vetoed, preventing the shower to
populate this region. At the end, the physical observables will
exhibit a dependence on the jet resolution parameter, $y_{cut}$, of
the next-to-next-to-leading log nature, i.e.
$\alpha_{s}^{k}log^{2k-2}y_{cut}$. We generated $q\bar{q}$ and
$q\bar{q}g$ events using the value for the jet resolution parameter of
0.0001 \cite{qcd}.  For higher $y_{cut}$ large discontinuities around
the cut value have been observed in the 2 $\to$ 3 jet rate
distribution as a function of $y_{23}$. The reason of their presence
is that SHERPA, being a tree level generator, cannot simulate
$q\bar{q}g$ events where one quark has very low energy or the two
quarks are very collinear, so such events were missing from the
simulated data sample. Such three jet $q\bar{q}g$ events, with a
highly energetic gluon and the other two quarks collinear, are largely
produced in the $J_{z}$ = 0 state since the $M_{\rm q}^{2}/s$
suppression is compensated by an $\alpha_{s}/s$ factor in the cross
section. Finally, the total cross sections given by SHERPA for the
$b\bar{b}(g)$ and $c\bar{c}(g)$ processes for the $J_{z}$ = 0 state
were scaled by a factor of 1.34 and 1.92 respectively, as one can see
in Figure \ref{sh-nlo}.  These K-factors resulted from a comparison
between the SHERPA cross sections and the theoretical NLO
calculations.

A total of 1000K events were generated for each
background process and each $\gamma \gamma$ spin state.

A convolution with the luminosity distribution is performed and a kinematic
cut of $\sqrt s_{\gamma \gamma}$ greater than 80 GeV
is imposed during the event generation for both signal and background processes.

The response of the detector has been simulated with
SIMDET 4 \cite{simdet}, a parametric Monte Carlo for the TESLA
$\EE$ detector. It includes a tracking and calorimeter simulation
and a reconstruction of energy-flow-objects (EFO).
Only EFOs with a polar angle above 7$^{0}$ can be taken
for the Higgs reconstruction simulating the acceptance
of the photon collider detector as the only deference
to the $\EE$ detector \cite{ggdet}.

The hadronic cross-section for $\gamma \gamma \to$hadrons events,
within the energy range above 2 GeV, is several hundred nb \cite{cs},
so that about 1.0 event of this type is produced per bunch crossing.
These events (pile-up) are overlayed to the signal events. Since the
pile-up events are produced in the t-channel $q$-exchange most of the
resulting final state particles are distributed at low angles.

\section{Event selection}
       
An intermediate mass Higgs production leads mainly to the final state:
$\gamma \gamma \to H \to b\bar{b}$. The major characteristics of these
events, used to distinguish the signal from the background, are the
event topology and the richness in b quarks. The background consists
of multi-jet events coming from $\gamma \gamma \to q\bar{q}(q)$
processes.

In order to minimise the pile-up contribution to the high energy
signal tracks the first step in the separation procedure was to reject
pile-up tracks as much as possible. The measurement of the impact
parameter of a particle along the beam axis with respect to the
primary vertex is used for this purpose, as described in Ref.
\cite{tgc}. A reconstruction of the angle of each EFO with respect to
the $z$-axis, $\theta_{EFO}$ makes it possible to distinguish further
between signal and pile-up EFOs. EFOs are rejected if $|\cos
(\theta_{EFO})|>$0.950.

\begin{figure*}[t]
\centering
\includegraphics[width=65mm]{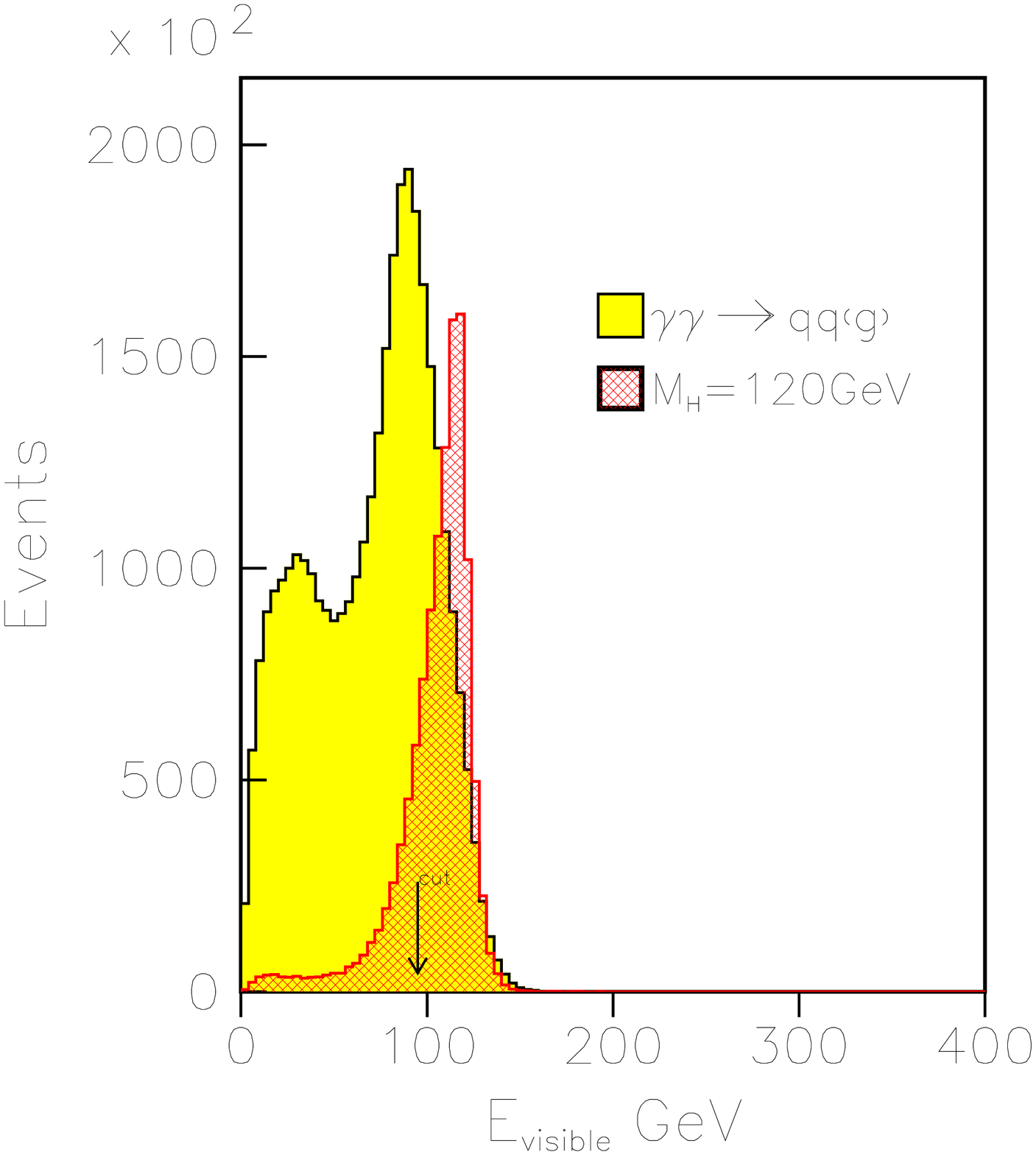}
\includegraphics[width=65mm]{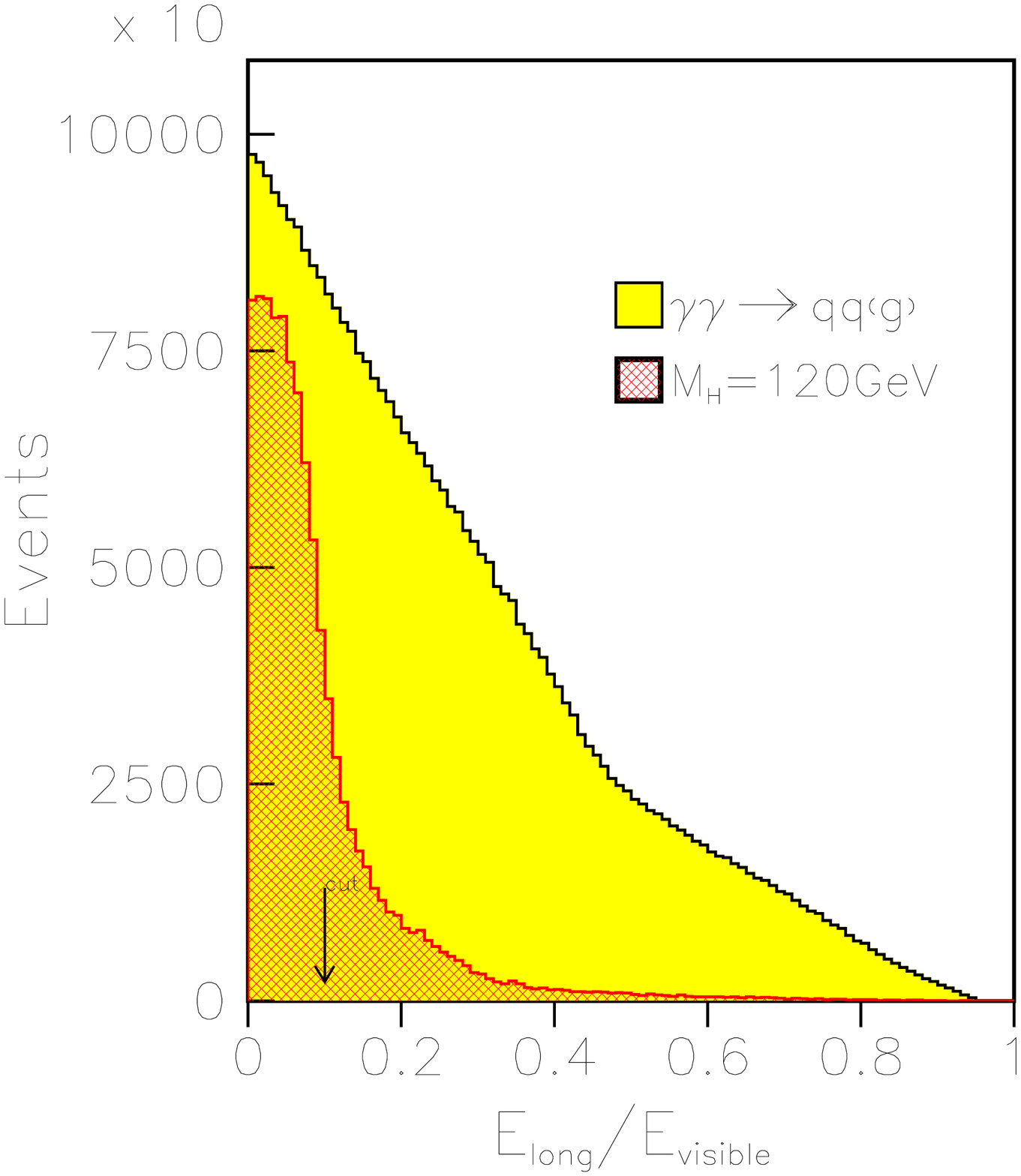} \\
\caption{Left: Distributions of the visible energy and Right: of the
  longitudinal imbalance for signal ($M_{H}$ = 120 GeV) and
  background. The distribution for the signal is arbitrarily
  normalised. Events with pile-up.  }
\label{fig_kine}
\end{figure*}

Hadronic balanced events are then selected requiring: large particle
multiplicity (at least 5 EFO), large visible energy ($E_{visible}$
greater than 95 GeV) and small longitudinal imbalance, normalised to
the visible energy (not larger than 0.1). Figure \ref{fig_kine} shows
the distributions of the visible energy and the longitudinal
imbalance.

\begin{figure*}[t]
\centering
\includegraphics[width=65mm]{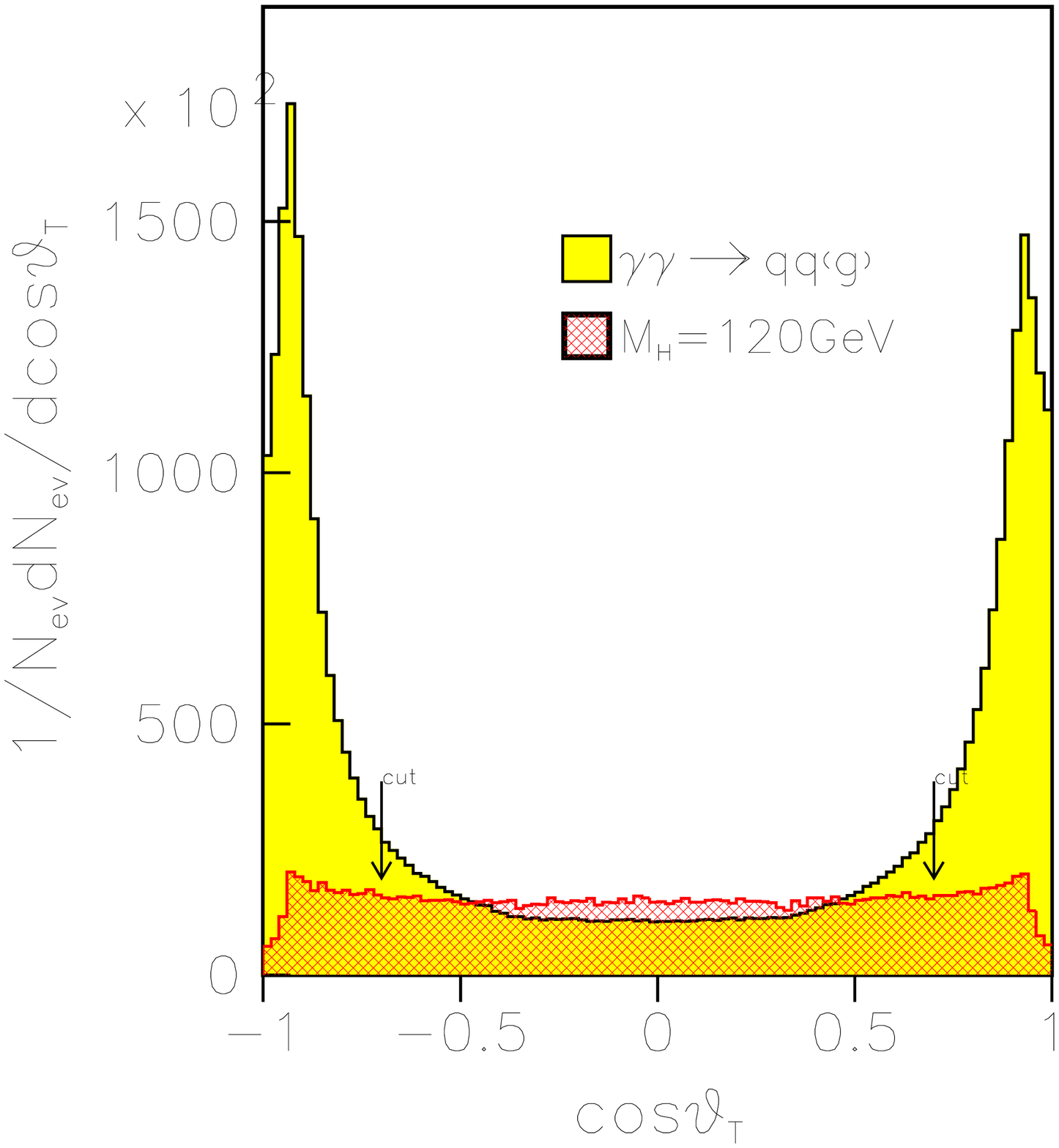}
\includegraphics[width=65mm]{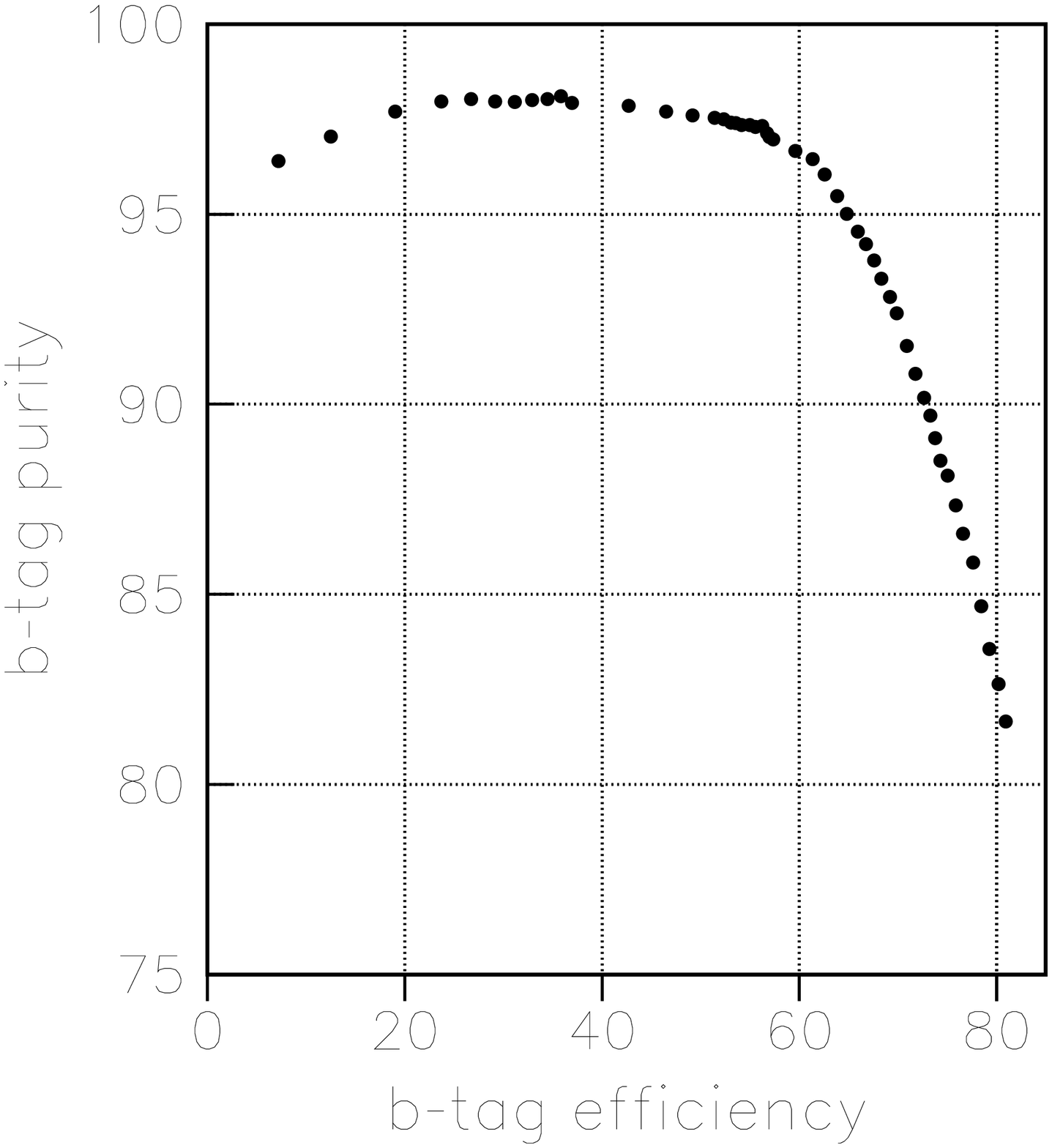} \\
\caption{Left: Distribution of the absolute value of the cosine of the
  thrust angle for signal ($M_{H}$ = 120 GeV) and background. The
  distribution for signal is arbitrarily normalised. Events with
  pile-up.  Right: Efficiency on b quarks and b-purity of the
  b-tagging for simulated $q \bar{q}$ events at $\sqrt{s}$ = $M_{Z}$.
}
\label{fig_thrust}
\end{figure*}

Due to the fact that the Higgs is centrally produced, the requirement
that the thrust of the event, see Figure \ref{fig_thrust} left ,
points in the central region of the detector ($|\cos \theta_{thrust}|
\le 0.7$) allows to reduce further the background while keeping a
large fraction of the signal.

In the remaining event sample
jets are reconstructed using the DURHAM clustering scheme \cite{durham}
with the resolution parameter y$_{\rm cut}$ = 0.02. Events are kept only
if there are at least 2 such jets.

The cross section for the continuum production of the charm quark is 16
times
larger than for bottom quarks. 
Therefore one of the most critical issues for this analysis is the 
capability of the detector to identify events in which a b quark is 
produced. To this aim a b-tagging algorithm based on a Neural Network has been 
applied. The algorithm combines several discriminating variables,
as for example, the impact parameter joint probability
introduced by ALEPH \cite{aleph} and
the $p_{t}$ corrected vertex invariant mass obtained with
the ZVTOP algorithm written for the SLD experiment \cite{zvtop}
into a feed forward Neural Network with 12 inputs and 3 
output nodes, described in Ref. \cite{nn}.

Figure \ref{fig_thrust} right shows the efficiency on b-quarks and the
b-quark purity for the algorithm exploited.  It has been obtained on a
Monte Carlo sample of $q\bar{q}$ events at $\sqrt{s}$ = $M_{Z}$.  The
b-tagging efficiency corresponding to a purity of 97$\%$ is 50$\%$.

\begin{figure*}[t]
\centering
\includegraphics[width=65mm]{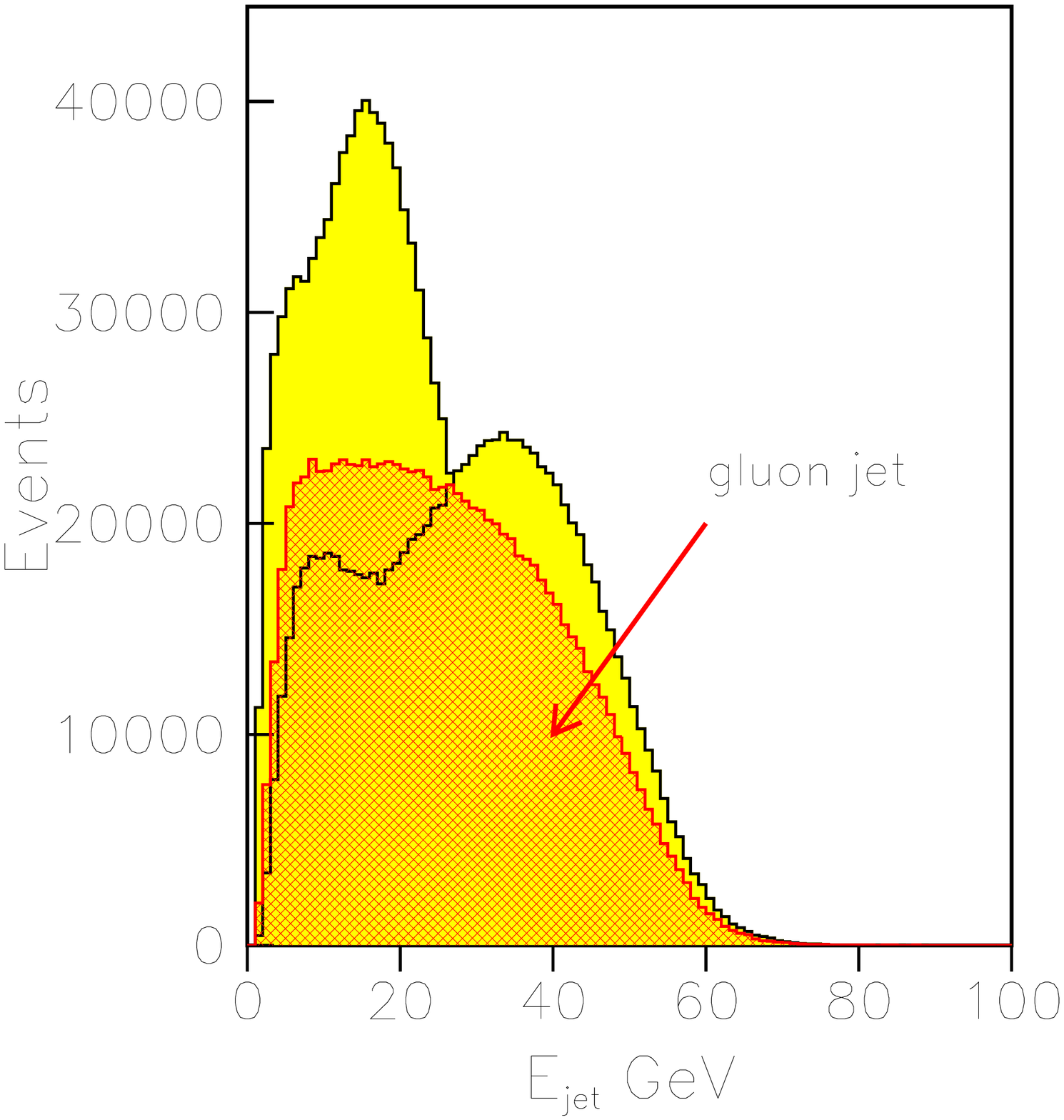}
\includegraphics[width=65mm]{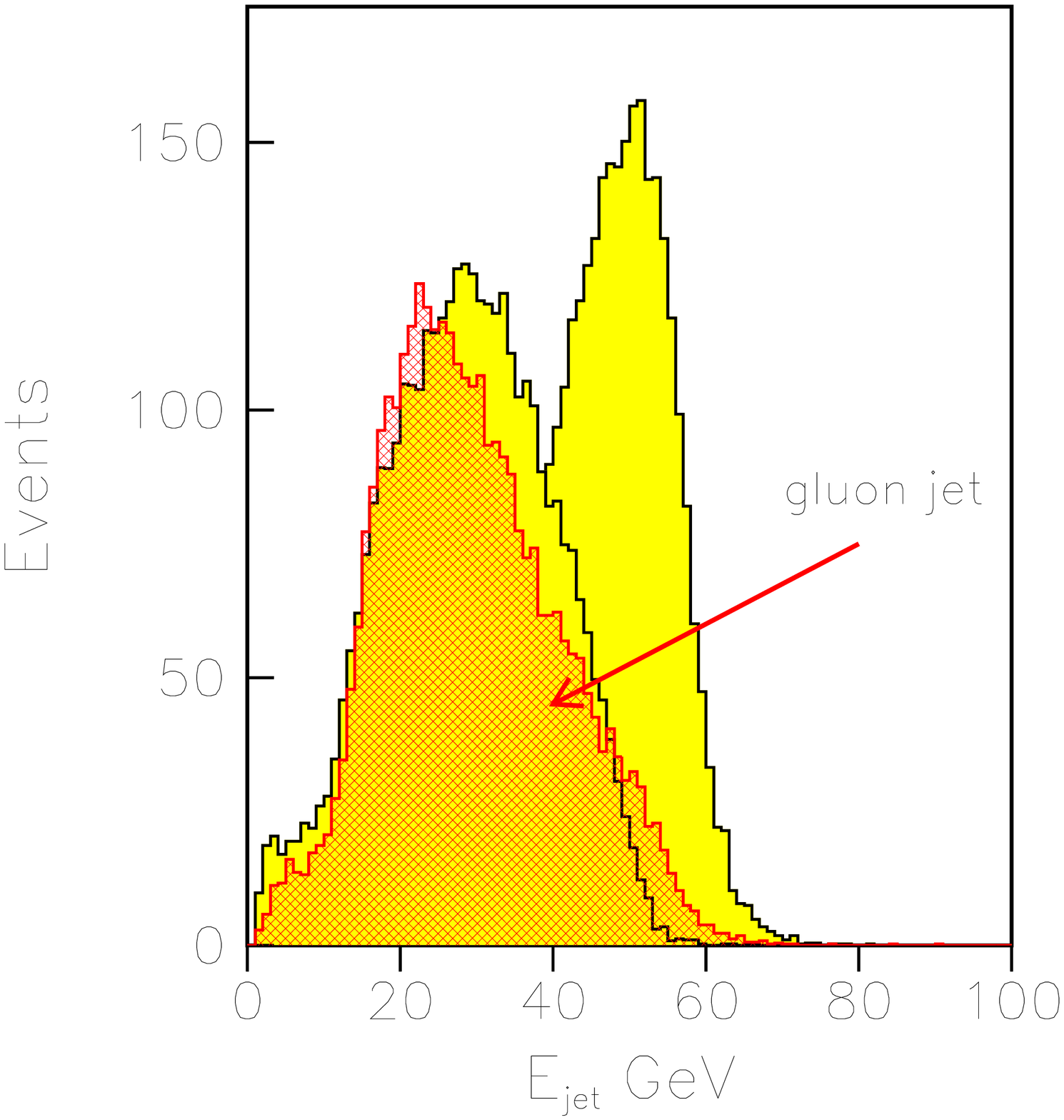} \\
\caption{Distributions of the b-quarks and gluon jets for the
  background (left) and the signal (right). Events with pile-up.}
\label{fig_energ}
\end{figure*}

The b-quarks coming from the decay of the Higgs boson are highly
energetic, whereas in the case of the background processes the gluon
and one b-quark jet are the most energetic. This is shown for 3-jet
events in Figure \ref{fig_energ}.  In order to reduce the background
further we look at the two fastest jets in the event and require the
${\rm NN}_{\rm out}$ to be greater than 0.9 for one jet and greater than 0.1
for the second one.  This procedure is also efficient for 2-jet
events. There is a large number of 2-jet background events where one b
is low energetic or both b-quarks are collinear so that they get merged
into one jet. For this reason 40\% of the J=0 2-jet events are
rejected by the b-tagging cut on the second jet while only 15\% of the
signal events fail this cut.

The total signal efficiency is estimated to be 22\% in the presence of 
the pile-up events.

\section{Results}

\begin{figure*}[t]
\centering
\includegraphics[width=0.48\linewidth]{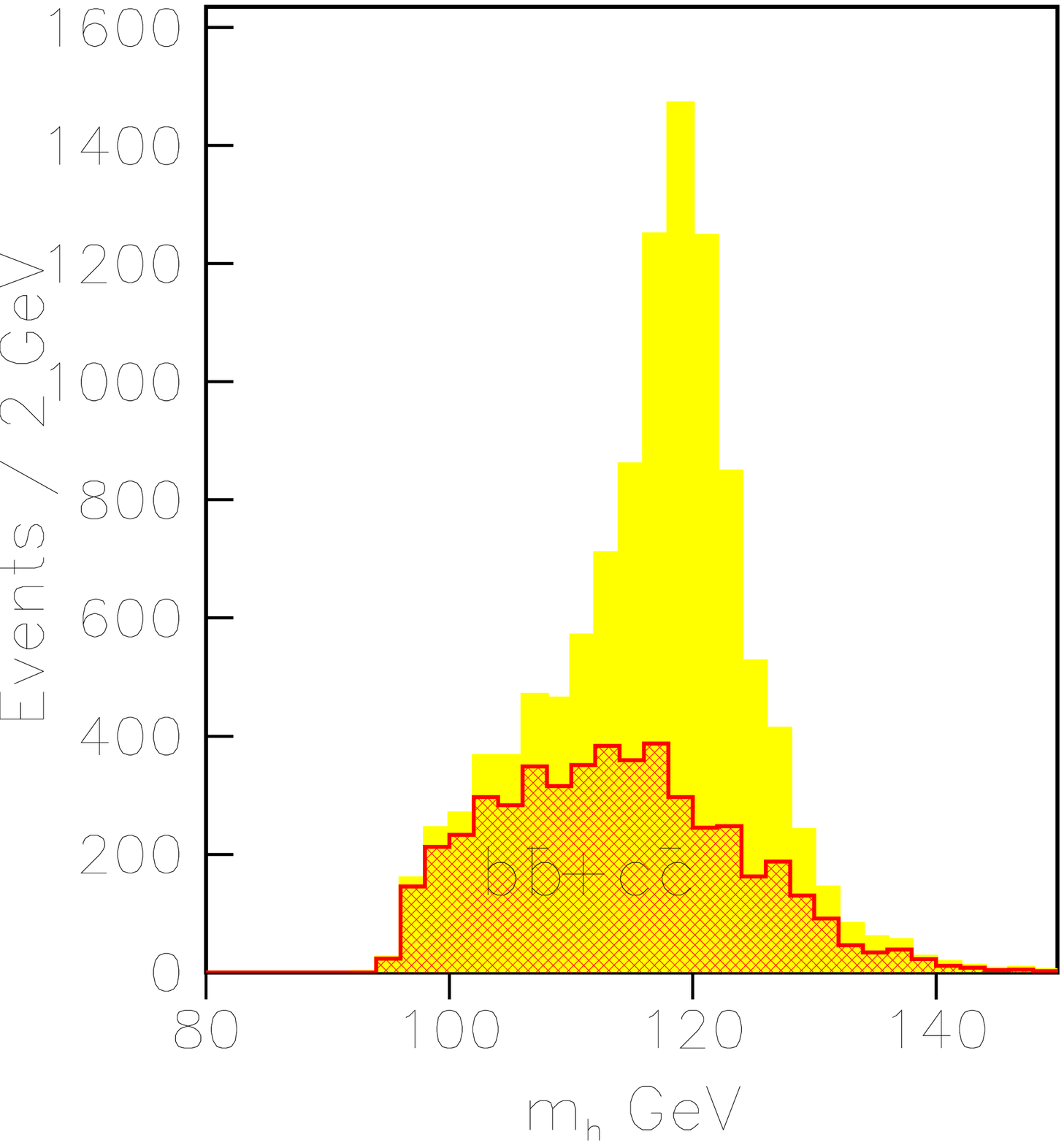} 
\includegraphics[width=0.48\linewidth]{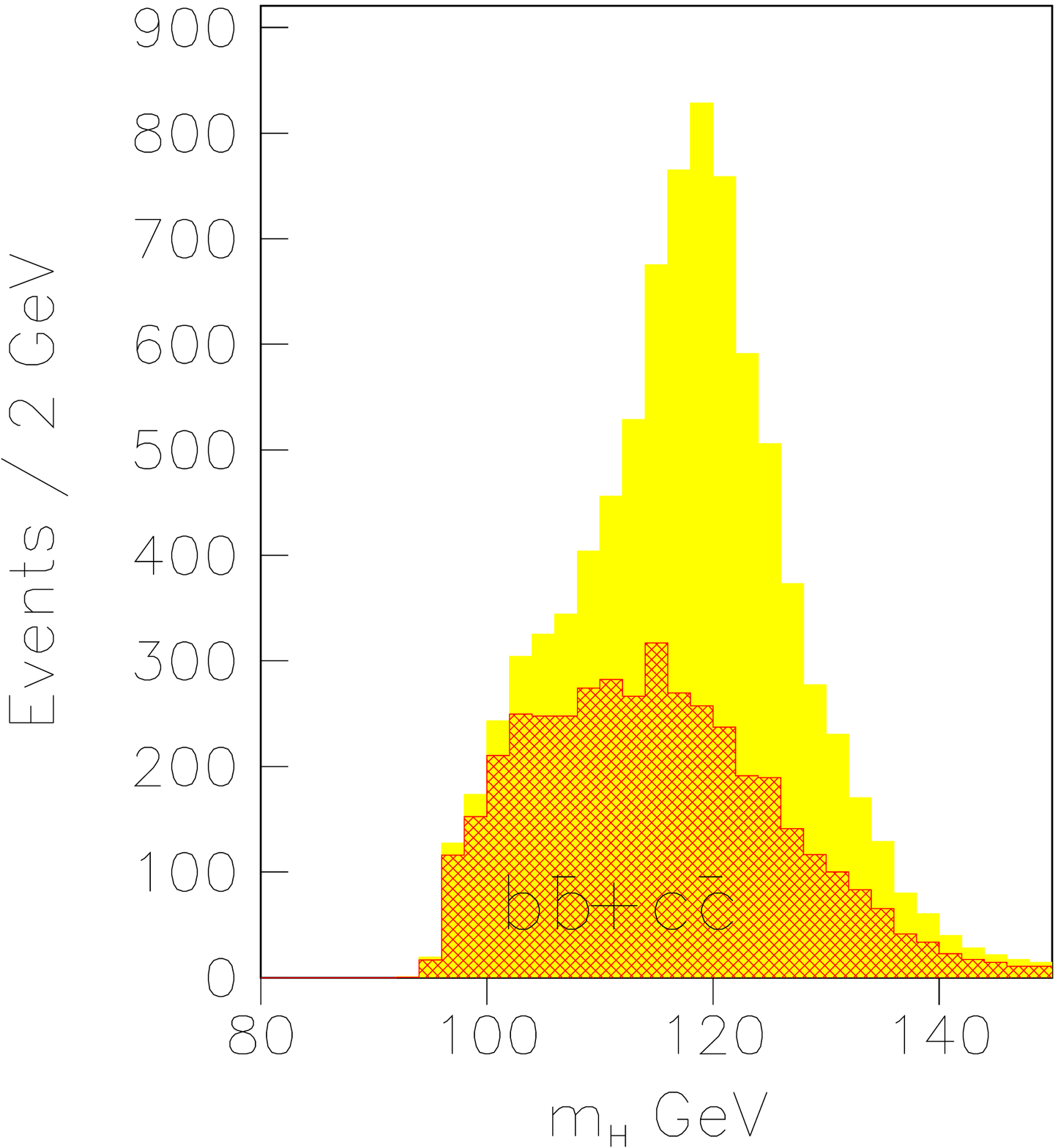}
\caption{Higgs invariant mass reconstruction on signal and background
  for a Higgs mass of 120 GeV without (left) and with (right) pile-up
  events overlayed.}
\label{fig_mass}
\end{figure*}

The reconstructed invariant mass for the selected signal and background
events
is shown in Figure \ref{fig_mass}. Here the invariant mass is corrected for
escaping neutrinos as in Ref. \cite{notes1}.
To enhance the signal a cut on the invariant mass is
tuned such that the statistical significance of the signal over background
is maximised. Events in the mass region of 112 GeV $ < M_{jets} < $ 134 GeV
are selected. The number of estimated signal and background events
in this window are 3534 and 2170, respectively.

The two photon decay width of the Higgs boson is
proportional to the event
rates of the Higgs signal. The statistical error of the number of signal
events,
$\sqrt N_{\rm obs}/({\it N} _{\rm obs}-{\it N}_{\rm bkg})$,
corresponds to the statistical error of this measurement. Here $N_{\rm
obs}$ is the
number of observed events, while $N_{\rm bkg}$ is the number of expected
background
events.

We obtain
$$\frac{\Delta [\Gamma (\rm H \to \gamma \gamma)\times \rm BR (\rm H \to
\rm b
\bar{\rm b})]}{[\Gamma (\rm H \to \gamma \gamma) \times \rm BR (\rm H \to
\rm b
\bar{\rm b})]}=\sqrt N_{\rm obs}/({\it N} _{\rm obs}-{\it N}_{\rm
bkg})=2.1\% .$$

\section{Conclusions}

The photon collider option at the ILC offers the possibility to
measure the partial width of the Higgs into photons, $\Gamma (\rm H
\to \gamma \gamma)$.  Taking higher order QCD corrections for the
background into account and using realistic assumptions for the
detector and background from pileup events We conclude that for a
Higgs boson with a mass $M_{\rm H}$ = 120 GeV $\Gamma (\rm H \to
\gamma \gamma) \times \rm BR (\rm H \to \rm b \bar{\rm b})$ can be
measured to 2.1$\%$. Using $\Delta \rm BR (\rm H \to \rm b \bar{\rm
  b})$ = 2 - 3$\%$ from the $e^+e^-$ mode of the ILC \cite{tdr} the
photonic width of the Higgs can be determined to 3$\%$.  At this
accuracy one can distinguish between the Standard Model Higgs particle
and the lightest scalar Higgs boson predicted by models beyond the
Standard Model. Also, the precise measurement of the decay width
$\Gamma (\rm H \to \gamma \gamma)$ can reveal heavy charged particles
circulating in the loop, as for example supersymmetric particles. The
accuracy of the mass determination of the heavier stop $\tilde t_{2}$
is estimated to be 10 - 20 GeV in \cite{stop2}, assuming that the
lighter stop $\tilde t_{1}$ and the mixing angle $\theta_{\tilde t}$
are known.

\section*{Acknowledgments}
The authors would like to thank Georgi Jikia, Frank Krauss and Andreas
Sch\"alicke for many interesting discussions. Part of this work was
supported by the CEEX Program of the Romanian 
Ministry of Education, Research and Youth, contract 05-D11-81/21.10.2005.

\end{document}